\documentclass[11pt,onecolumn,floatfix,altaffilletter,superscriptaddress,tightenlines,showpacs,showkeys,preprintnumbers,nofootinbib]{revtex4-2}
\pdfoutput=1
\usepackage[colorlinks=true,citecolor=blue,linkcolor=blue,breaklinks=true]{hyperref}
\usepackage{amsmath,amssymb,empheq}
\usepackage{epsfig} 
\usepackage{graphicx,wasysym}        
\usepackage{url}
\usepackage{color}
\usepackage{multirow}
\usepackage{placeins}
\usepackage[dvipsnames]{xcolor}
\usepackage{braket}
\usepackage{float}
\usepackage{slashed}
\usepackage{comment}
\usepackage{fdsymbol}
\hypersetup{
  colorlinks=true,
  linkcolor=red,
  filecolor=magenta,   
  urlcolor=blue,
  citecolor=blue,
}

\hypersetup{colorlinks,linkcolor={Blue},citecolor={teal},urlcolor={violet}}

\allowdisplaybreaks

\bibpunct{[}{]}{,}{n}{}{,}

\def\beq{\begin{equation}}
\def\eeq{\end{equation}}
\def\bea{\begin{eqnarray}}
\def\eea{\end{eqnarray}}
\makeatletter
\@addtoreset{equation}{section}

\newcommand{\INFN}{INFN - Sezione di Napoli, Complesso Universitario Monte S. Angelo, I-80126 Napoli, Italy}

\newcommand{\SSM}{Scuola Superiore Meridionale, Università degli studi di Napoli ``Federico II'', Largo San Marcellino 10, 80138 Napoli, Italy}
\newcommand{\NNU}{Institute of Theoretical Physics and Institute of Physics Frontiers and Interdisciplinary Sciences, Nanjing Normal University, Wenyuan Road, Nanjing, Jiangsu, 210023, China}
\newcommand{\MEKL}{Nanjing Key Laboratory of Particle Physics and Astrophysics, Nanjing, 210023, China}
\begin{document}
\title{Multifaceted Supercooling: From PTA to LIGO}
    \author{Satyabrata Datta}
   \email{amisatyabrata703@gmail.com}
   \affiliation{\NNU}
   \affiliation{\MEKL}
   \author{Rome Samanta}
  \email{samanta@na.infn.it}
  \affiliation{\SSM}
  \affiliation{\INFN}
  
\begin{abstract}
 
Supercooled phase transitions, as predicted, e.g., in near-conformal and confining extensions of the Standard Model (SM), are established sources of strong stochastic gravitational wave backgrounds (SGWBs). In this work, we investigate another facet of such transitions: their significant and largely uncharted impact on gravitational wave spectra originating from independent cosmological sources. Focusing on gravitational waves produced by a metastable cosmic string network, we show that an intervening supercooled phase, initiating thermal inflation, can reshape and suppress the high-frequency part of the spectrum. This mechanism reopens regions of string parameter space previously excluded by LIGO’s null results, while remaining compatible with the nanohertz SGWB signal reported by pulsar timing arrays (PTAs). The resulting total spectrum typically exhibits a dual-component structure, sourced by both string decay and the phase transition itself, rendering the scenario observationally distinctive. We systematically classify the viable parameter space and identify regions accessible to upcoming detectors such as Advanced LIGO, LISA, and ET.
\end{abstract}
\maketitle
\newpage
\tableofcontents
\section{Introduction}

Strong first-order phase transitions (FOPT) in the early universe are of significant interest in both particle physics and cosmology due to their rich phenomenology and potential observational signatures, see, e.g., \cite{Caprini:2018mtu,Cap1,Cap2,Kamionkowski:1993fg,Apreda:2001us,Kosowsky:1992vn,Grojean:2006bp,Huber:2008hg,Ashoorioon:2009nf,Hind1,Hind2,Hind3,Hind4,Hind5,Rop1,Rop2,Rop3,Megevand:2016lpr,Tenkanen:2016idg,Megias:2018sxv,Croon:2018erz,Creminelli:2001th,Randall:2006py,Konstandin:2011dr,vonHarling:2017yew,Iso:2017uuu,Ellis:2018mja,Ellis:2020nnr,Ferrer:2023uwz,Jinno:2016knw,Marzola:2017jzl,Jinno:2019jhi,Nardini:2007me,DelleRose:2019pgi,Baratella:2018pxi,VonHarling:2019rgb,Kanemura:2024pae,Liu:2024fly,Athron:2022mmm,Athron:2023mer,Athron:2023xlk,Athron:2023aqe,Athron:2019teq,Borah:2022cdx,Goncalves:2024vkj,Conaci:2024tlc,Arteaga:2024vde,DiBari:2021dri,DiBari:2020bvn,Banerjee:2024fam,Banerjee:2023qya}. In particular, supercooled first-order phase transitions (scFOPT) \cite{Megevand:2016lpr,Ellis:2018mja}, where the transition is delayed well below the critical temperature due to a substantial barrier between metastable and true vacua, can lead to highly non-equilibrium dynamics with far-reaching implications. In scenarios exhibiting significant supercooling, the universe remains trapped in a metastable vacuum for an extended period, during which vacuum energy dominates the energy density \cite{Guth:1983kw,Ellis:2018mja,Athron:2022mmm,Athron:2023mer,Athron:2023xlk}. Once nucleation becomes efficient, the transition proceeds rapidly via the nucleation and percolation of true vacuum bubbles. The associated dynamics—bubble wall collisions, bulk plasma motion, and turbulence—can serve as a powerful source of a stochastic gravitational wave (GW) background. These GWs, if within the sensitivity range of current or future detectors such as advanced LIGO run (LVK-Design) \cite{LIGOScientific:2014pky}, LISA \cite{lisa}, DECIGO \cite{decigo}, and ET \cite{et}, provide a unique observational window into high-scale physics beyond the Standard Model (BSM)\cite{Caprini:2018mtu,Cap1,Cap2}.

On the other side, the vacuum energy domination during the supercooled phase transition can temporarily induce a phase of thermal inflation \cite{Guth:1983kw,Ellis:2018mja}. Unlike standard inflationary models, thermal inflation is a low-scale, short-duration epoch that can efficiently dilute pre-existing relics such as moduli, gravitinos, or topological defects. The reheating following bubble collisions restores radiation domination and sets the stage for the subsequent thermal history required for successful Big Bang Nucleosynthesis (BBN). In this work, we investigate this thermally inflating phase induced by supercooling and analyze its impact on a metastable cosmic string network that forms prior to the onset of vacuum domination.

\begin{figure}
    \centering
    \includegraphics[scale=1.2]{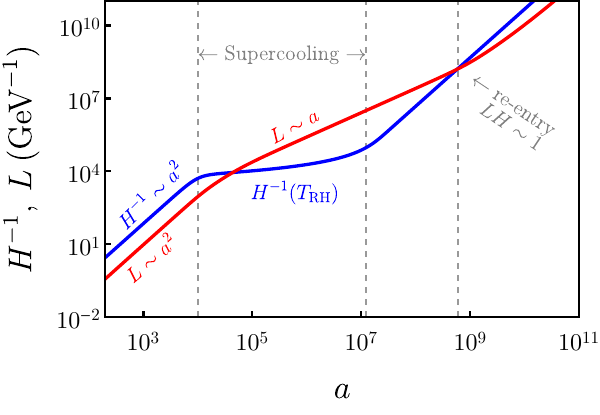}
   \caption{
After formation, the string network enters a scaling regime with $L \sim a^2$, where $L$ is the long-string correlation length and $a$ is the scale factor. During supercooling, the rapid expansion stretches the network beyond the horizon, halting loop production, and the correlation length scales as $L \sim a$. After supercooling, as the universe reheats, the strings gradually re-enter the horizon and the network returns to scaling with $L \sim a^2$. This figure is generated using Eqs.~\eqref{rad}--\eqref{avsz}, with the scale factor normalized to unity at $z = z_{\rm in}$. We maintain this normalization in all subsequent figures where $a$ is shown.
}
    \label{fig1}
\end{figure}

The GW spectrum sourced by a cosmic string network~\cite{Vilenkin:1981bx,Vachaspati:1984gt,Hill:1987qx} provides a powerful diagnostic of the universe’s thermal history \cite{Cui:2018rwi,Gouttenoire:2019kij,Guedes:2018afo,Cui:2019kkd,Gouttenoire:2019rtn,Samanta:2021zzk,Borah:2022byb,Datta:2025yow}. Deviations from the standard radiation-dominated evolution, such as early matter domination (EMD) or transient inflationary phases, can leave distinctive imprints on the high-frequency plateau of the spectrum. These modifications typically result in a red-tilted feature at high frequencies, arising from the suppression of GW emission during non-standard epochs. In particular, during a phase of vacuum energy domination, the rapid accelerated expansion stretches the long-string correlation length beyond the Hubble horizon \cite{Guedes:2018afo,Cui:2019kkd}, significantly reducing the production of loops—the primary emitters of gravitational radiation.  Consequently, GW emission is suppressed during this period, and a significant contribution resumes only after the network re-enters the horizon following the end of vacuum domination (see Fig. \ref{fig1}). This transition introduces a characteristic spectral break frequency, above which the GW spectrum falls, reflecting the imprint of the non-standard cosmological era.

While such high-frequency suppression in the GW spectrum due to non-standard expansion is a generic feature of any cosmic string network, metastable string networks \cite{meta1,meta2,meta3,meta4,meta5,meta6,meta7,meta8,Pallis:2024mip,Hu:2025sxv,Chitose:2023dam,Chitose:2024pmz,Maji:2025thf} are of particular phenomenological interest. In such a case, the resulting GW spectrum can simultaneously account for the stochastic gravitational wave background (SGWB) observed by pulsar timing array (PTA) experiments at nanohertz frequencies~\cite{ng1,ng2,ng3,ng4,ng5,EPTA:2023xxk}, while evading current upper bounds from ground-based interferometers such as LIGO in the hertz regime \cite{KAGRA:2021kbb}. Moreover, this scenario yields robust and distinctive predictions for next-generation detectors operating at intermediate to high frequencies, including LVK-D, LISA, DECIGO, and ET.

This work is structured around two central phenomenological objectives. First, focusing on the region of metastable string parameter space favored by PTA observations, we identify the corresponding parameters of a supercooled phase transition that induces a period of vacuum domination, responsible for the spectral suppression at high frequencies--thus ensuring consistency with the absence of a signal in the LIGO band. Second, we examine additional GW contributions sourced directly by the phase transition within the same parameter space. Inclusion of this component introduces spectral features that are sufficiently distinct to differentiate the supercooled scenario from an EMD-induced suppression, despite both mechanisms producing similar effects on the metastable cosmic string–sourced spectrum \cite{Antusch:2024ypp,Datta:2024bqp}. Both objectives are successfully met, positioning supercooling as a compelling and testable scenario for multi-frequency GW observations, with amplitudes anchored by PTA data.

The remainder of this paper is organized as follows. In Section~\ref{s2}, we present the general setup of the scenario. Section~\ref{s3} contains our numerical results and highlights how the impact of supercooling on a metastable cosmic string network differs from the more commonly studied EMD case. Finally, we conclude in Section~\ref{s4}.

\section{Imprints of supercooling on GWs from metastable strings}\label{s2}

Gravitational waves are emitted by cosmic string loops that are chopped off from long strings formed during the spontaneous breaking of the gauged $U(1)$ symmetry~\cite{Vilenkin:1981bx,Vachaspati:1984gt}. The long string network is characterized by a correlation length 
$L = \sqrt{\mu / \rho_\infty}$, where $\rho_\infty$ is the energy density in long strings, and $\mu$ is the string tension defined as~\cite{Hill:1987qx}
\begin{equation}
\mu = \pi v_\Phi^2\, h(\lambda, g'), \label{tension}
\end{equation}
with $h(\lambda, g') \simeq 1$ unless the scalar self-coupling $\lambda$ and gauge coupling $g'$ are strongly hierarchical~\cite{sh1,sh2}.

The time evolution of a radiating loop with initial size $l_i = \alpha t_i$ is given by
\[
l(t) = l_i - \Gamma G \mu (t - t_i),
\]
where $\Gamma \simeq 50$~\cite{Vilenkin:1981bx,Vachaspati:1984gt}, $\alpha \simeq 0.1$~\cite{Blanco-Pillado:2013qja,Blanco-Pillado:2017oxo}, $G$ is the Newton's constant, and $t_i$ is the time of loop formation.

The total energy emitted in GWs can be decomposed into harmonics with instantaneous frequencies 
\[
f_k = \frac{2k}{l_k} = \frac{a(t_0)}{a(t)} f,
\]
where $k = 1, 2, 3, \ldots, k_{\rm max}$, $f$ is the observed frequency today at $t_0$, and $a(t)$ is the scale factor. The total GW energy density is then given by summing over all modes~\cite{Blanco-Pillado:2013qja,Blanco-Pillado:2017oxo}:
\begin{equation}
\Omega_{\rm GW}(f) = \sum_{k=1}^{k_{\rm max}} \frac{2k \mathcal{F}_\alpha G \mu^2 \Gamma_k}{f \rho_c \alpha (\alpha + \Gamma G \mu)} \int_{t_F}^{t_0} \frac{A_\beta(t_i)}{t_i^4} \left[\frac{a(t)}{a(t_0)}\right]^5 \left[\frac{a(t_i)}{a(t)}\right]^3  \Theta(t_i - t_{\rm fric}) \Theta\left(t_i - \frac{l_c}{\alpha}\right) dt. \label{gwcs1}
\end{equation}
Here, $\mathcal{F}_\alpha \simeq 0.1$ is an efficiency factor~\cite{Blanco-Pillado:2013qja}, and $A_\beta(t_i)$ is the loop formation efficiency, computable from the velocity-dependent one-scale model~\cite{Martins:1996jp,Martins:2000cs,Sousa:2013aaa,Auclair:2019wcv,Sousa:2020sxs}. The quantity $\Gamma_k={\Gamma k^{-\delta}}/{\zeta(\delta)}$ quantifies the emitted power in $k$-th mode, with $\delta=4/3$ ($\delta=5/3$) for loops containing cusps (kinks) \cite{Damour:2001bk}, making it evident that  $k=1$ (fundamental) mode dominantly contribute to the GWs. The integration is performed over the GW emission time $t$, with $t_i$ marking the loop formation time and $t_F$ representing the time when the network forms. The Heaviside functions apply cutoffs: (i) loops radiate GWs once the network overcomes friction at $t_{\rm fric}$~\cite{Vilenkin:1991zk}; (ii) particle production is subdominant to the gravitational radiation for $l\gtrsim l_c$~\cite{Matsunami:2019fss,Auclair:2019jip}. Both cutoffs introduce a spectral break at $f_*$, above which the gravitational wave spectrum for the fundamental mode of loop oscillation falls off as $f^{-1}$. This high-frequency suppression mimics that of intermediate non-standard cosmologies. However, since our focus is on strong-amplitude gravitational waves, corresponding to large string tensions favored by PTA observations, these high-frequency constraints are less relevant for the present discussion \cite{Datta:2024bqp}.

Metastable string networks evolve similarly to stable ones initially, but eventually decay via Schwinger pair production of monopole-antimonopole pairs through quantum tunnelling. These pairs act as endpoints, fragmenting the strings into finite segments. The decay rate per unit length in the semiclassical approximation is given by
\begin{equation}
\Gamma_d = \frac{\mu}{2\pi} \exp(-\pi \kappa), \qquad \sqrt{\kappa} = \frac{m_M}{\sqrt{\mu}} \simeq \frac{\Lambda_{\rm GUT}}{v_\Phi}, \label{meta}
\end{equation}
where $m_M$ is the monopole mass, approximately $\Lambda_{\rm GUT} \simeq 10^{16}$ GeV. From Eq.~\eqref{meta}, one can define the decay time 
$t_s = \frac{1}{\sqrt{\Gamma_d}}\gg t_i$, before which the network behaves like a stable string network. After $t_s$, the network decays, and no new loops are produced. For $t > t_s$, the loop number density receives an exponential suppression~\cite{meta6}:
\begin{equation}
D(l, t) = \exp\left[-\Gamma_d \left( l(t - t_s) + \frac{1}{2} \Gamma G \mu (t - t_s)^2 \right) \right] \Theta\left( \alpha t_s - \bar{l}(t_s) \right),
\end{equation}
where $\bar{l}(t_s) \simeq l + \Gamma G \mu t$. The Heaviside function ensures that only loops formed before $t_s$ contribute to the density.

Unless $\kappa$ is large (which stabilizes string network until today), the upper limit of integration in Eq.~\eqref{gwcs1} can be taken as $t \sim t_{\rm eq}$, because for $t_s < t < t_{\rm eq}$ the loop number density is exponentially suppressed. In this work, we focus on GW emission from loops. However, if monopoles do not carry unconfined flux, string segments formed by monopole nucleation may also contribute to the GW background~\cite{meta6}. A few typical features emerge from the integration in Eq.~\eqref{gwcs1}: an infrared tail that scales as $f^2$—a power law favored by PTA observations, a plateau at intermediate frequencies, and finally a spectral break at high frequencies. This break can arise either from the Heaviside functions present in Eq.~\eqref{gwcs1} or from a transient non-standard cosmological phase, such as the supercooling scenario considered in this work. In the case of supercooling, the effect is equivalent to performing the integral with the instantaneous value of $A_\beta$ and by including an additional Heaviside function enforcing $t_i > t_{\rm re}$, where $t_{\rm re}$ denotes the time at which the cosmic string network re-enters the horizon following the stretching caused by the supercooling phase. A more accurate description can be obtained by solving all the dynamical quantities numerically and then fitting the relevant observables with the numerical results. While all the relevant quantities can be solved with any variable, the scale factor $a$, the time $t$, and the temperature $T$, we consider $z=T_{\rm RH}/T$, with $T_{\rm RH}$ being the reheating temperature, as the variable to be consistent with the notation in our previous publication on metastable strings plus EMD by primordial black holes \cite{Datta:2024bqp}. In this case, the following six coupled equations need to be solved and fed into Eq.\eqref{gwcs1}: 

\begin{empheq}[box=\fbox, left=\empheqlbrace]{align}
\frac{d\rho_R}{dz} &= -\frac{4}{z}\rho_R,\label{rad} \\
\frac{d\rho_V}{dz} &= \Gamma_V \frac{1}{z\tilde{\mathcal{H}}} \rho_V, \\
\frac{dt}{dz} &= \frac{1}{\tilde{\mathcal{H}} z}, \\
\frac{d\xi}{dz} &= \frac{1}{t \tilde{\mathcal{H}} z} \left( H t \xi (1 + \tilde{v})^2 + \frac{1}{2} \tilde{c} \tilde{v} - \xi \right), \\
\frac{d\tilde{v}}{dz} &= \frac{1 - \tilde{v}^2}{\tilde{\mathcal{H}} z} \left( \frac{k(\tilde{v})}{\xi t} - 2 H \tilde{v} \right), \\
\frac{da}{dz} &= \left(1 - \frac{\mathcal{K}}{\tilde{\mathcal{H}}} \right) \frac{a}{z} \label{avsz},
\end{empheq}
Here, $\rho_i$ represent the energy densities of radiation ($\rho_R$) and vacuum ($\rho_V$), and $H$ denotes the Hubble parameter. The vacuum decay rate is parametrized as $\Gamma_V \simeq H(z_{\rm RH})/N_e$, where $N_e$ is the number of e-folds during the supercooling phase. The quantity $\mathcal{K}$ accounts for the energy transfer between components and is defined as $\mathcal{K} = -\Gamma_V \rho_V / (4 \rho_R)$, while the effective Hubble-like parameter is given by $\tilde{\mathcal{H}} = H + \mathcal{K}$.
\begin{figure}
\centering 
\includegraphics[scale=.75]{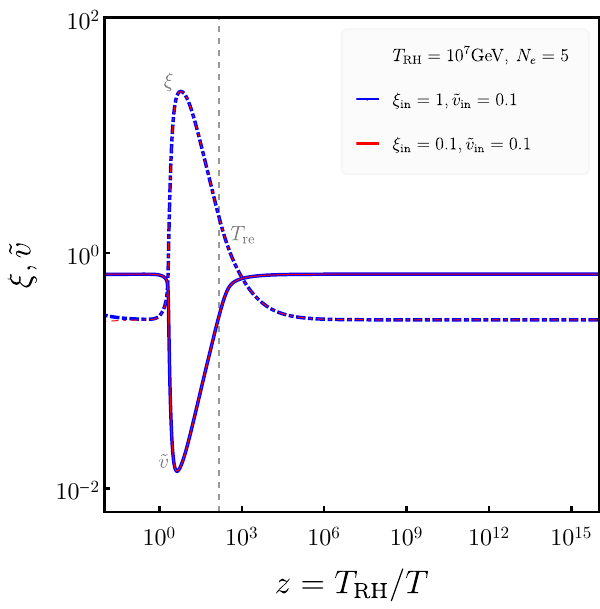} 
\includegraphics[scale=.75]{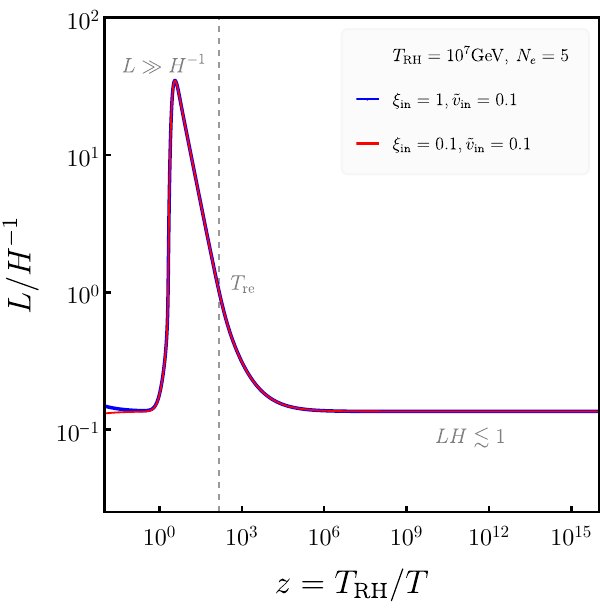} \\
\includegraphics[scale=.75]{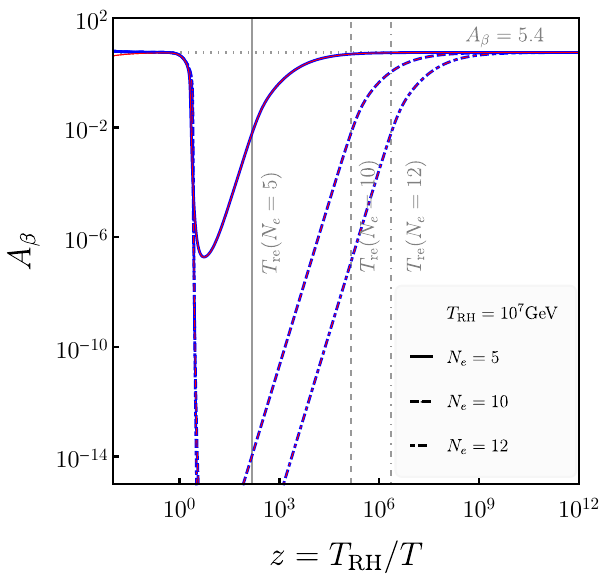} 
\includegraphics[scale=.75]{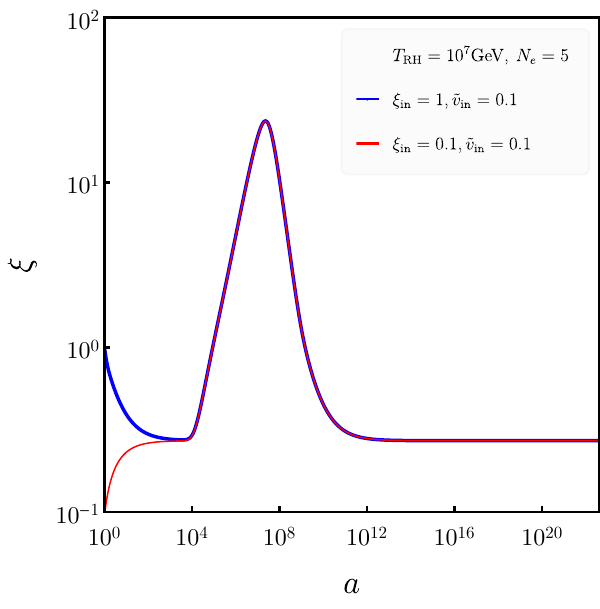}
\caption{
\textit{Top Left}: Evolution of $\xi$ and  $\tilde{v}$.  \textit{Top Right}: Evolution of the long string correlation length normalized to the Hubble horizon. \textit{Bottom Left}: Evolution of the loop production efficiency $A_\beta$ for different numbers of e-folds during the supercooled phase.  \textit{Bottom Right}: Evolution of $\xi$ and $\tilde{v}$ as functions of the scale factor $a$.
}
\label{fig2}
\end{figure}
We assume an instantaneous reheating scenario\footnote{If the decay rate of the scalar field is suppressed, i.e.,  $\Gamma_\Phi\ll H(T_p)$, the scalar field can oscillate around the true vacuum for an extended period and can lead to a subsequent matter-like era \cite{Ellis:2019oqb}. Consequently, depending on the decay rate, the reentry of the string network may occur much later than in the efficient reheating scenario. We shall provide a comprehensive analysis on this in a follow-up work.}, with the entropy and energy relativistic degrees of freedom taken to be equal and constant, i.e., $g_{*s} = g_{*\rho} = \text{const}$\footnote{For a quantitative comparison with the exact degrees of freedom scenario \cite{Saikawa:2018rcs}, please see Appendix \ref{appA}.}. 

In the cosmic string sector, $\tilde{v}$ is the root-mean-square (RMS) velocity of the long string network, and $\tilde{c} = 0.23 \pm 0.04$ is a loop chopping efficiency parameter. The function $k(\tilde{v})$ phenomenologically incorporates the effects of small-scale structure on the strings and is given by
\begin{equation}
k(\tilde{v}) = \frac{2\sqrt{2}}{\pi}(1 - \tilde{v}^2)(1 + 2\sqrt{2} \tilde{v}^3) \frac{1 - 8 \tilde{v}^6}{1 + 8 \tilde{v}^6}.
\end{equation}

The characteristic length scale of the network is defined as $L = \xi(t)\, t$, where $\xi(t)$ is a dimensionless correlation length. The loop production efficiency, $A_\beta$, is expressed as
\begin{equation}
A_\beta = \frac{\tilde{c} \tilde{v}}{\sqrt{2} \, \xi^3}.
\end{equation}

\begin{figure}
\centering 
\includegraphics[scale=.75]{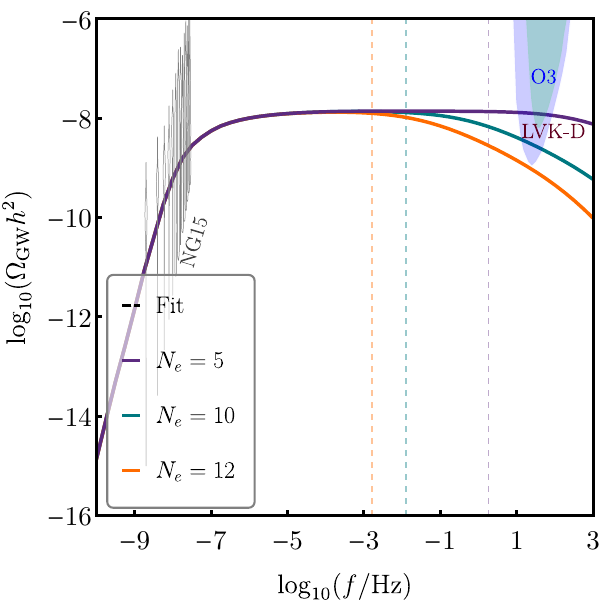} \includegraphics[scale=.75]{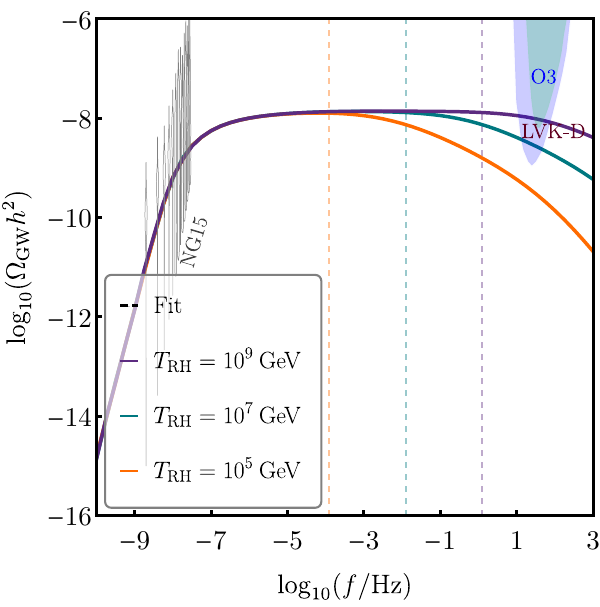}
\caption{
\textit{Left}: Gravitational wave spectra for $G\mu = 10^{-7}$, $\sqrt{\kappa} = 7.9$, and $T_{\rm RH} = 10^7~\mathrm{GeV}$, shown for different values of the number of e-folds $N_e$. 
\textit{Right}: Gravitational wave spectra for $G\mu = 10^{-7}$, $\sqrt{\kappa} = 7.9$, and fixed $N_e = 10$, plotted for varying reheating temperatures $T_{\rm RH}$. 
In both panels, the dashed vertical lines indicate the fitted spectral break corresponding to the analytically derived reheating temperature $T_{\rm re}$.
}
\label{fig3}
\end{figure}

In the top panel of Fig.~\ref{fig2}, we present the evolution of $\xi$ and $\tilde{v}$ (left), and the normalized characteristic length $L/H^{-1}$ (right), as functions of $z$, for two different initial conditions: $\tilde{v}_{\rm in} = 0.1$ with $\xi_{\rm in} = 0.1$ (red curves) and $\xi_{\rm in} = 1$ (blue curves). The bottom left panel shows the evolution of the loop production efficiency factor $A_\beta$ for different values of e-folding.

These plots collectively illustrate how the presence of a supercooling phase causes the string network to freeze, leading to a dramatic suppression in loop production and, consequently, a significant reduction in gravitational wave emission. The re-entry time of the network into the Hubble horizon, i.e., when $L H \lesssim 1$, depends sensitively on the duration of the supercooling phase, or equivalently, the number of e-folds. A longer supercooling phase results in a larger $L$, thus delaying the onset of thawing and the restoration of standard network evolution. Once the network re-enters the Hubble horizon, $\tilde{v}$ begins to increase from near-zero values, and $L$ evolves toward the scaling regime, which is typically assumed in gravitational wave computations in the literature. 

The sharp stretching of the network as a function of $z$ is a consequence of our assumption of efficient reheating, which implies the vacuum energy scale $T_V \sim T_{\rm RH}$. A less steep variation is observed when the same evolution is plotted against the scale factor, using Eq.~\eqref{avsz}, as shown in the bottom right panel for $\xi$ versus $a$.

In Fig.~\ref{fig3}, we present the GW spectra summing over $10^4$ modes for benchmark values of $T_{\rm RH}$ (left panel) and $N_e$ (right panel), demonstrating the suppression of the signal at high frequencies while remaining consistent with the amplitude indicated by PTA observations \footnote{Note that, unlike earlier results based solely on NANOGrav data which were consistent with stable cosmic strings \cite{csfit1,csfit2,csfit3,csfit4, Hindmarsh:2022awe, Kume:2024adn}, the latest data from multiple PTA collaborations, now in good agreement with each other, disfavor stable cosmic strings due to the observed flatness of the spectrum in the PTA frequency band. On the other hand, the metastable scenario fits well because the combined contribution from the stable and decaying loops 
provides a blue tilted infrared spectrum within the PTA band \cite{metafit2,metafit3,metafit4,metafit5,metafit6}.}. The horizontal lines correspond to a fitting formula for the spectral break frequency, derived in the Appendix \ref{appB}:

\begin{equation}
f_{\rm brk}^{\rm sc} = 3.9\times 10^{-6}~{\rm Hz} \left( \frac{g_*(T_{\rm RH})}{g_*(T_0)} \right)^{1/4} \left( \frac{0.1 \times 50 \times 10^{-6}}{\alpha \Gamma G\mu} \right)^{1/2} \left( \frac{T_{\rm RH}}{e^{N_e}} \right). \label{fbrk}
\end{equation}

This expression provides a useful analytical approximation to the spectral break predicted by the VOS model, making it well-suited for parameter space exploration (numerical scanning of the parameter space). With these foundations in place, we now proceed to a more detailed numerical analysis and a discussion on the distinctive features of the supercooling plus metastable string scenario.
 \section{Numerical analysis and Multifacetedness of supercooling }\label{s3}

To enable a more rigorous exploration of the parameter space, we extend our analysis beyond the spectral break frequency given in Eq.~\eqref{fbrk}. Specifically, when a large number of modes are summed, the GW spectrum is expected to scale as $f^{-1/3}$ at high frequencies \cite{Blasi:2020wpy,csfit4}. In this regime, the GW amplitude beyond the spectral break can be approximated as

\begin{equation}
\Omega_{\rm GW}(f) \simeq \Omega_{\rm GW}^{\rm plt,\infty} \left( \frac{f}{f_{\rm brk}^{\rm sc}} \right)^{-1/3} \Theta\left(f - f_{\rm brk}^{\rm sc} \right), \quad \text{with} \quad \Omega_{\rm GW}^{\rm plt,\infty} \simeq 3.1 \times 10^{-8} \left( \frac{G\mu}{10^{-7}} \right)^{1/2}, \label{gwpbh}
\end{equation}

where $\Omega_{\rm GW}^{\rm plt,\infty}$ represents the plateau amplitude before the spectral break. Using Eq.~\eqref{fbrk}, the above expression can be recast in the form

\begin{equation}
\Omega_{\rm GW}(f) \simeq \Omega_{\rm GW}^{\rm plt,\infty} \cdot \Omega_{\rm GW}^{\rm mod,sc},
\end{equation}

where the modulation factor $\Omega_{\rm GW}^{\rm mod,sc}$, accounting for the impact of a non-standard cosmological phase (e.g., intermediate inflation or supercooling), is given by

\begin{equation}
\Omega_{\rm GW}^{\rm mod,sc} \simeq 0.0115\, \left( \frac{g_*(T_{\rm RH})}{100} \right)^{1/12} \left( \frac{10^{-6}}{G\mu} \right)^{1/6} \left( \frac{e^{10}}{e^{N_e}} \right)^{1/3} \left( \frac{T_{\rm RH}}{10^5\, \mathrm{GeV}} \right)^{1/3} \left( \frac{f}{25\, \mathrm{Hz}} \right)^{-1/3}.
\end{equation}

In the absence of such non-standard epochs, $\Omega_{\rm GW}^{\rm mod,sc} = 1$. Under this assumption, the non-observation of a stochastic GW background by the LIGO O3 run—i.e., $\Omega_{\rm GW}^{\rm any\, model} \lesssim 1.7 \times 10^{-8}$ at $f_{\rm LIGO} = 25\,\mathrm{Hz}$—implies that string tensions $G\mu \gtrsim 3 \times 10^{-8}$ are excluded. Therefore, the string tension values favored by PTA data are already in tension with LIGO constraints, unless $\Omega_{\rm GW}^{\rm mod,sc}(f = 25\,\mathrm{Hz}) \ll 1$ due to suppression from a non-standard cosmological history. 

Taking into account the modulation factor, we now categorize the parameter space into the following three distinct regimes:

\begin{itemize}
    \item \textbf{Category A:} 
    \[
    \Omega_{\rm GW}^{\rm plt,\infty} \gtrsim \Omega_{\rm GW}^{\rm LIGO-O3}, \quad \text{and} \quad 
    \Omega_{\rm GW}^{\rm LIGO-O3} \gtrsim \Omega_{\rm GW}^{\rm plt,\infty} \cdot \Omega_{\rm GW}^{\rm mod,PBH}(f = 25\,\mathrm{Hz}) \gtrsim \Omega_{\rm GW}^{\rm LVK-D}.
    \]
    In this region, the unmodulated plateau amplitude exceeds current LIGO-O3 bounds, but the modulated spectrum remains within LIGO-O3 sensitivity and above the LVK design threshold.

    \item \textbf{Category B:} 
    \[
    \Omega_{\rm GW}^{\rm LIGO-O3} \gtrsim \Omega_{\rm GW}^{\rm plt,\infty} \gtrsim \Omega_{\rm GW}^{\rm LVK-D}, \quad \text{and} \quad 
    \Omega_{\rm GW}^{\rm LIGO-O3} \gtrsim \Omega_{\rm GW}^{\rm plt,\infty} \cdot \Omega_{\rm GW}^{\rm mod,PBH}(f = 25\,\mathrm{Hz}) \gtrsim \Omega_{\rm GW}^{\rm LVK-D}.
    \]
    This region is consistent with both current and projected sensitivities: the plateau amplitude lies below current bounds but above LVK design reach, and remains viable after modulation.

    \item \textbf{Category C:} 
    \[
    \Omega_{\rm GW}^{\rm plt,\infty} \cdot \Omega_{\rm GW}^{\rm mod,PBH}(f = 25\,\mathrm{Hz}) \lesssim \Omega_{\rm GW}^{\rm LVK-D}, \quad \text{for all values of} \quad \Omega_{\rm GW}^{\rm plt,\infty}.
    \]
    Here, the modulated GW signal is too suppressed to be detectable even at LVK design sensitivity. Detectors operating in the same frequency ballpark, such as ET and CE, can probe this category.
\end{itemize}
\begin{figure}
    \centering
    \includegraphics[scale=0.58]{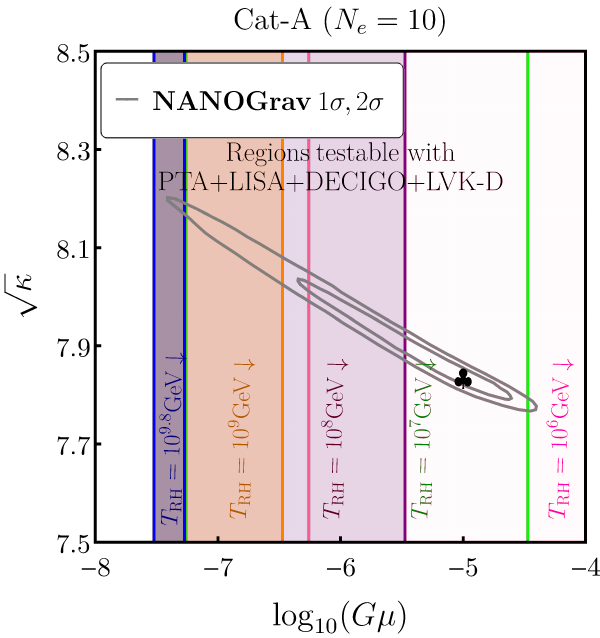}\includegraphics[scale=0.53]{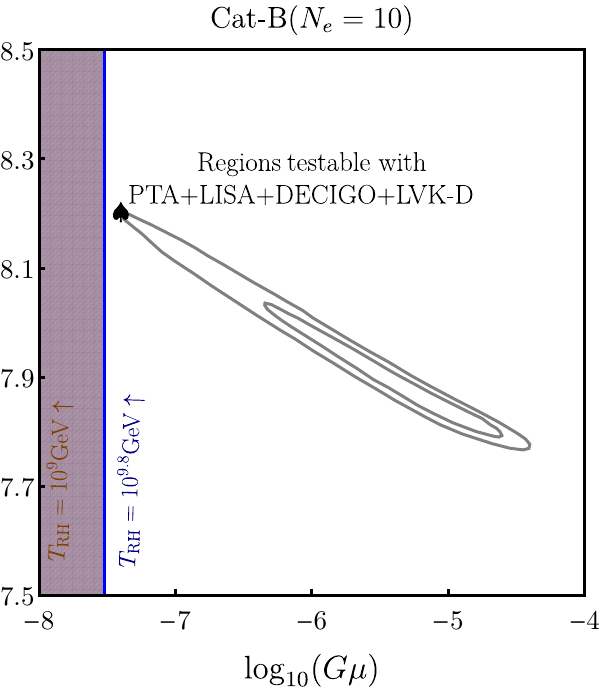}\includegraphics[scale=0.53]{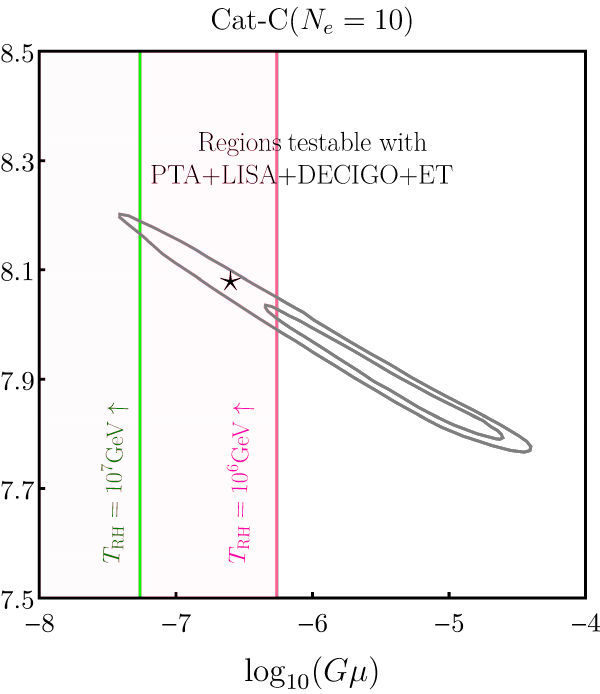}\\
  \caption{
\textit{Left}: Colored regions on the $G\mu$–$\sqrt{\kappa}$ plane, labeled by different values of $T_{\rm RH}$, correspond to gravitational wave spectra with plateau amplitudes $\Omega_{\rm GW}^{\rm plt,\infty} \gtrsim \Omega_{\rm GW}^{\rm LIGO-O3} \equiv 1.7 \times 10^{-8}$ and are therefore testable with LVK-Design sensitivity at $25~\mathrm{Hz}$. In the absence of supercooling, these regions would be excluded by LIGO-O3, since the spectra do not fall off sufficiently at $f \ll 25~\mathrm{Hz}$ and exceed the observational bound. Gray contours indicate regions consistent with the NANOGrav 15-year data.  \textit{Middle}: Colored regions where $\Omega_{\rm GW}^{\rm LIGO-O3} \gtrsim \Omega_{\rm GW}^{\rm plt,\infty} \gtrsim \Omega_{\rm GW}^{\rm LVK-D} \equiv 2 \times 10^{-9}$. Although in principle testable by LVK-D, these parameter regions are already disfavored by PTA observations even without invoking supercooling, and thus are less relevant to our discussion. \textit{Right}: Regions yielding lower GW amplitudes at LIGO frequencies, and therefore not accessible to LVK-D, but within reach of future detectors such as the ET. Benchmark points selected from each category are illustrated in the gravitational wave spectra shown in Fig.~\ref{fig5} (left).
}

    \label{fig4}
\end{figure}
\noindent
In the above, $\Omega_{\rm GW}^{\rm LIGO-O3}$ is the upper bound from the LIGO-O3 run, and $\Omega_{\rm GW}^{\rm LVK-D}$ denotes the minimum detectable amplitude projected by the LVK design sensitivity. In Fig.~\ref{fig4}, we show the regions of parameter space that are ``rescued" in each category for $N_e = 10$ and various values of $T_{\rm RH}$. The black contours represent the $1\sigma$ and $2\sigma$ confidence regions favored by NANOGrav—one of the PTA collaborations \cite{ng1,ng5}. While we focus on NANOGrav due to its higher statistical significance, we note that a global analysis using other PTA data sets yields qualitatively similar results \cite{InternationalPulsarTimingArray:2023mzf}.

Categories A and C emerge as complementary regions in the parameter space, while Category B appears to be entirely disfavored by the PTA data. This outcome is consistent with our earlier observation that the LIGO O3 constraint nearly rules out the parameter space associated with metastable strings. However, this conclusion slightly overestimates the exclusion because, in our analysis, we have neglected the variation of the effective degrees of freedom $g_*$ in the computation of the GW spectrum. Including this effect would allow for a small portion of the parameter space to be recovered.

 \begin{figure}
\includegraphics[scale=.75]{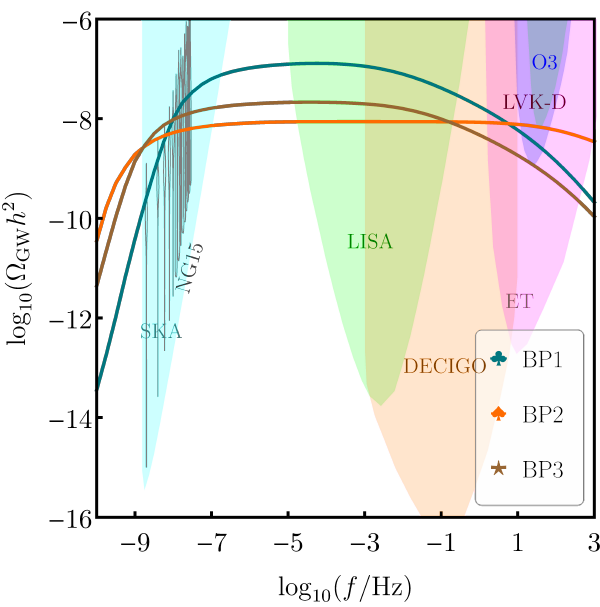} \includegraphics[scale=.75]{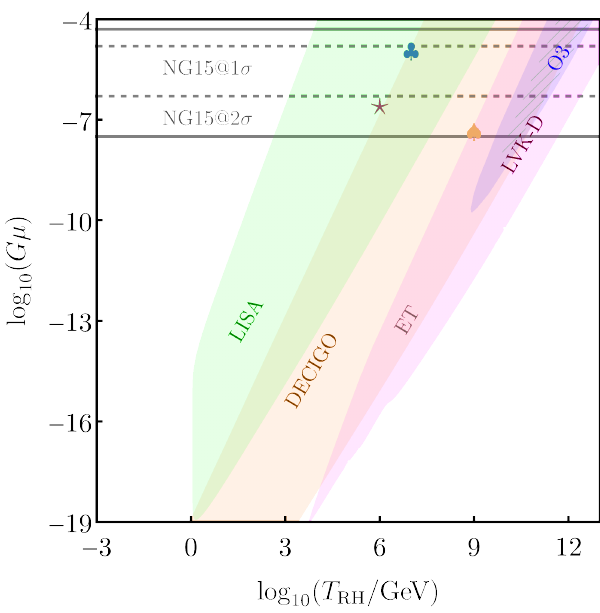}

 	\caption{
\textit{Left}: Representative GW spectra corresponding to the benchmark points selected from each category shown in Fig.~\ref{fig4}. \textit{Right}: Sensitivity of current and future GW detectors to the spectral break frequency. The NANOGrav 15-year confidence regions are indicated by the horizontal solid and dashed lines. The benchmark points from the left panel are also marked for reference.
}
\label{fig5}
 \end{figure}
In the left panel of Fig.~\ref{fig5}, we show the GW spectra corresponding to each of the three categories discussed earlier, with representative benchmark points marked as $\clubsuit$, $\spadesuit$, and $\star$ in Fig.~\ref{fig4}. An additional point of interest is the potential of future GW detectors to observe the spectral break frequency induced by supercooling. While our primary focus remains on the PTA-favored region of the string parameter space, we also explore lower-amplitude GW signals and their associated spectral breaks resulting from supercooling dynamics. The results of this extended scan are presented in the right panel of Fig.~\ref{fig5}, where we highlight the regions that could be probed by future missions: LISA, DECIGO, and ET. 

An interesting outcome of our analysis is that, due to the PTA preference for relatively large values of $G\mu$, the resulting spectral break induced by the vacuum-dominated epoch falls within the observable frequency range even for high reheating temperatures $T_{\rm RH}$. This stands in contrast to the case of stable cosmic strings, where large $G\mu$ values are strongly constrained by PTA data, pushing the corresponding spectral features outside the accessible range of current detectors.
 It is important to note, however, that a definitive identification of the physical origin of a detected spectral break would require a statistical analysis, such as parameter reconstruction or Bayesian inference \cite{Caprini:2019pxz}, as similar features could arise from a variety of alternative scenarios discussed earlier. In Appendix~\ref{appB}, we provide a comparative discussion of the EMD and supercooling scenarios. While both scenarios can yield similar spectral shapes, we highlight that the underlying model parameters, such as the reheating temperature $T_{\rm RH}$, can differ significantly.

One of the central motivations of this work is to emphasize that the supercooling scenario, distinct both conceptually and dynamically from other early-universe phases such as EMD, can be clearly distinguished by its capacity to generate additional sources of gravitational waves. Importantly, we identify a sizable region of parameter space where the resulting combined spectral features are characteristic of supercooling and cannot be easily mimicked by alternative cosmological histories.

  \begin{figure}
\centering
\includegraphics[scale=.75]{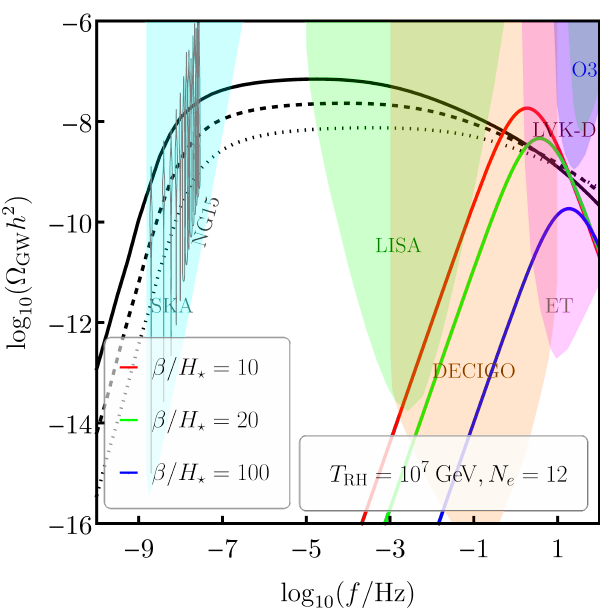}
\includegraphics[scale=.75]{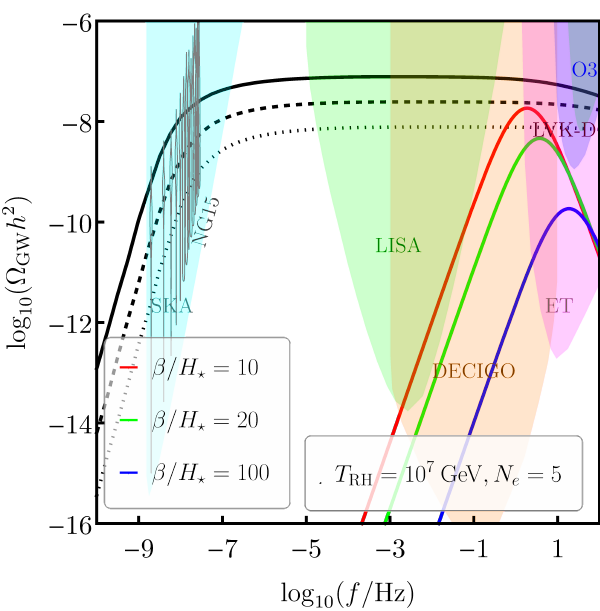}
 	\caption{
\textit{Left}: Gravitational wave spectra for a long-duration supercooling scenario, in which the FOPT signal dominates over the suppressed contribution from cosmic strings. This configuration is particularly distinctive and cannot be easily mimicked by alternative nonstandard cosmologies, such as an EMD era. \textit{Right}: Spectrum corresponding to a shorter-duration supercooling scenario, where the cosmic string-induced plateau dominates. In this case, potential degeneracies with EMD scenarios may arise, making it less distinguishable based solely on spectral features.
}
\label{fig6}
 \end{figure}
Among the most extensively studied sources of GWs at the particle-cosmology interface are first-order phase transitions, which can give rise to GWs through: (1) bubble wall collisions, (2) sound waves in the relativistic plasma, and (3) turbulence or magnetohydrodynamic (MHD) effects. In the case of strong supercooling, the resulting GW spectral shape is particularly similar to that arising from bubble wall collisions. To model this, we adopt the following commonly used ansatz~\cite{Lewicki:2020azd,Lewicki:2022pdb}, which has been also used to study the effect of supercooling on GW spectra from stable cosmic strings \cite{Ferrer:2023uwz}:

\begin{equation}
    \Omega_{\rm GW}^{\rm PT}(f)\, h^2 = 1.67 \times 10^{-5} \left( \frac{100}{g_*(T_{\rm RH})} \right)^{1/3} \left( \frac{\beta}{H_\star} \right)^{-2} \left( \frac{\alpha_{\rm PT}}{1 + \alpha_{\rm PT}} \right)^2 
    \frac{A(a + b)^c}{\left[ b \left( \frac{f}{f_{\rm peak}} \right)^{-a/c} + a \left( \frac{f}{f_{\rm peak}} \right)^{b/c} \right]},
\end{equation}

where the shape parameters are taken to be $a = b = 2.4$, $c = 4.0$, $A = 5.13 \times 10^{-2}$, and $\alpha_{\rm PT}=e^{4 N_e}\gg1$. The parameter $\beta/H_\star$ representing the inverse timescale of the transition defines the GW spectral profile along with  $T_{\rm RH}$. The peak frequency of the spectrum is given by

\begin{equation}
    f_{\rm peak} = 1.65 \times 10^{-5}~\text{Hz} \left( \frac{T_{\rm RH}}{100~\text{GeV}} \right) 
    \left( \frac{g_*}{100} \right)^{1/6} \left( \frac{0.71}{2\pi} \right) \left( \frac{\beta}{H_\star} \right).
\end{equation}

   \begin{figure}
\centering
\includegraphics[scale=.75]{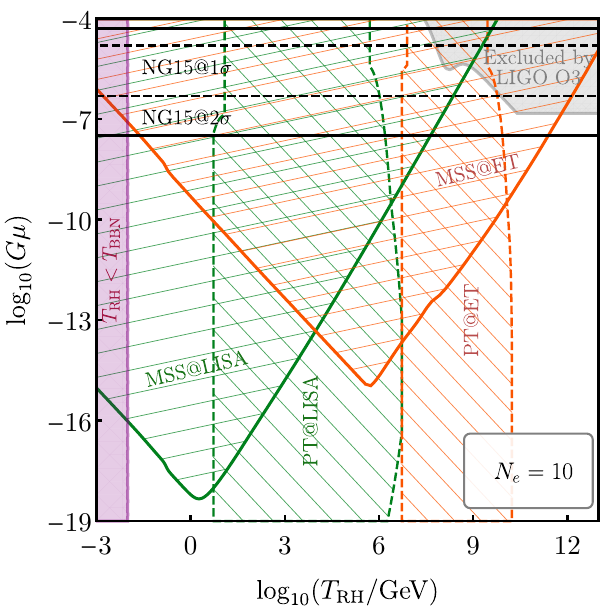}
\includegraphics[scale=.75]{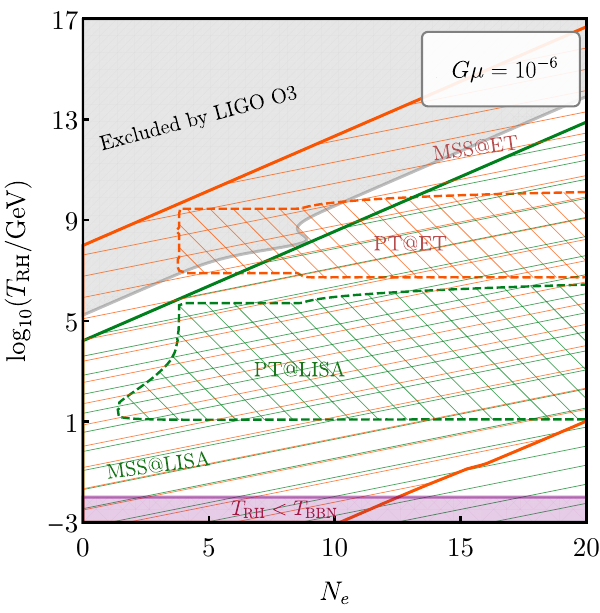}
\caption{
\textit{Left}: Region of parameter space where the modified gravitational wave signal, featuring both cosmic string and FOPT contributions, is detectable with signal-to-noise ratio $\mathrm{SNR} \gtrsim 10$. The intersection of the CS and FOPT regions represents the ``smoking gun" parameter space where FOPT dominates. The analysis separates the modified cosmic string tail (feature beyond $f_{\rm brk}^{\rm sc}$) from the FOPT signal, considering only portions of the spectrum where the amplitude deviates by at least 10\% from that of a standard string spectrum with the same $G\mu$. For this plot, we fix the phase transition parameter at $\beta/H_\star = 10$ and the duration of supercooling at $N_e = 10$. \textit{Right}: Similar parameter scan, but with $G\mu$ held fixed while varying the duration of supercooling $N_e$.
}
\label{fig7}
 \end{figure}
In Fig.~\ref{fig6}, we present two representative plots that highlight the GW signals originating from supercooled phase transitions and those from cosmic strings. The left panel illustrates a scenario in which the GW signal from a strongly supercooled phase transition dominates over the suppressed cosmic string background, thereby offering a clear distinction from alternative early-universe histories. On the other hand, the right panel shows a case with a small number of e-folds, corresponding to a brief period of vacuum energy domination, where the suppression of the cosmic string GW amplitude is relatively mild. In such cases, the resulting GW spectrum may appear degenerate with that expected from an EMD phase, complicating efforts to reconstruct the underlying cosmological history using GW data alone.

To explore this further, Fig.~\ref{fig7} provides a detailed analysis of the parameter space where the phase transition GW signal dominates over the cosmic string contribution. Here we show the detectability of the modified smoking gun signal coming from the combined GW contributions from strings and FOPT by evaluating the corresponding signal-to-noise ratios (SNR) expected in various GW detectors. We have chosen a smaller value for $\beta/H_\star=10$ to highlight this unique peak feature at high frequencies. In most of the parameter space with higher $G\mu$, the string contribution will be the dominant one at lower frequencies. For the GW spectrum from PT, we focused only on regions where it is practically detectable above the background from the strings. We considered only the modified parts of the spectrum where the amplitude deviates by at least $10\%$ away from the standard string spectrum for a given value of $G\mu$. The overlap of these two regions indicates the combined ability to detect the unique modified signals. The left panel shows a generic scan over the parameter space, while the right panel focuses on a benchmark value of $G\mu$ favored by PTA observations, highlighting the observational prospects for distinguishing these scenarios.

We conclude by highlighting several additional aspects of the supercooling scenario that merit attention. First, for sufficiently small values of $\beta/H_\star$, scalar-induced gravitational waves (SIGWs) sourced by curvature perturbations during the supercooled first-order phase transition (scFOPT) can produce a dominant low-frequency GW component~\cite{Lewicki:2024sfw}. Second, and notably, within the PTA-favored parameter space (interpreting the data as GWs from metastable strings), it is also feasible to produce primordial black holes (PBHs) constituting up to 100\% of the dark matter abundance. For readers interested in these extended phenomenological implications, concise discussions on both SIGW production and PBH formation are provided in Appendix~\ref{appc1}. 

Finally, the overall framework discussed here benefits from a reasonable degree of flexibility between two key parameters: the string tension and the reheating temperature following supercooling. This allows one to simultaneously fit PTA data using metastable strings while suppressing the high-frequency GW signal below the LIGO band. Although we do not present a complete model in this article, a compelling realization of this scenario can arise in multistep hierarchical models discussed such as in Refs.~\cite {Fornal:2020esl,DiBari:2023mwu}.

In frameworks where a common phase transition generates both cosmic strings and GWs from a FOPT, the spectral features typically (assuming reasonable signal strength from both sources) show up at high frequencies due to the intrinsic correlation between the string tension and the reheating scale (see, e.g., Eq.\eqref{cor}). For a quantitative discussion of such a unified scenario in the context of a minimal conformal $B\!-\!L$ model, we refer the reader to Appendix~\ref{appc2}.

\section{conclusion}\label{s4}
In this work, we have investigated the distinctive gravitational wave (GW) signatures arising from a supercooled first-order phase transition (scFOPT) in the context of metastable cosmic strings. Our analysis focused on the impact of a transient vacuum energy domination on the evolution of the string network. Using the velocity-dependent one-scale (VOS) model, we demonstrated how Hubble-induced stretching during the supercooling phase significantly suppresses loop production, thereby reducing the GW amplitude at high frequencies. This suppression is strongly dependent on the duration of supercooling, parameterized by the number of e-folds $N_e$ and the reheating temperature $T_{\rm RH}$.

A central motivation for this study is the recent evidence from pulsar timing arrays (PTAs), which may suggest a stochastic gravitational wave background (SGWB) originating from cosmic strings. However, interpreting these signals in terms of metastable cosmic strings is in significant tension with upper limits from the LIGO-O3 run, particularly at higher frequencies. We investigated whether this tension can be resolved by introducing a period of supercooling, which naturally suppresses the high-frequency portion of the string-induced GW spectrum. By systematically categorizing the parameter space based on the detectability of this high-frequency tail, we found that although the standard string scenario is largely excluded, the presence of supercooling reopens the full PTA-preferred region. Moreover, it offers new opportunities to probe the suppressed GW spectrum with future detectors such as LVK at design sensitivity (LVK-D) and the Einstein Telescope (ET).

Beyond the string contribution, we examined the possibility that the same phase transition responsible for supercooling could also produce an additional GW component. In the strongly supercooled regime, this signal can dominate over the suppressed cosmic string-induced GW background, leading to a characteristic multi-sourced GW spectrum. On the other hand, scenarios with only brief vacuum domination (i.e., small $N_e$) exhibit minimal suppression of the string spectrum and may be degenerate with early matter domination (EMD), complicating the identification of the underlying cosmological history. We also analyzed the prospects for observing these features with future detectors such as LISA, DECIGO, and the Einstein Telescope (ET), emphasizing their sensitivity to the break frequency as a key discriminator. 

In summary, our results highlight the power of gravitational wave astronomy to probe not only cosmic strings but also the thermal and inflationary history of the early universe. The supercooled metastable string scenario offers a compelling and testable framework with rich phenomenology, capable of explaining PTA observations while remaining compatible with high-frequency constraints. Its unique spectral features make it a promising target for current and next-generation GW experiments.

\section*{Acknowledgements}
 The work of SD is supported by the National Natural Science Foundation of China (NNSFC) under grant No. 12150610460. The work of  RS is supported by the research project TAsP (Theoretical Astroparticle Physics) funded by the Istituto Nazionale di Fisica Nucleare (INFN).

\appendix
 \section{Constant vs exact relativistic degrees of freedom $g_*(T)$} \label{appA}
\begin{figure}
\centering
\includegraphics[scale=.75]{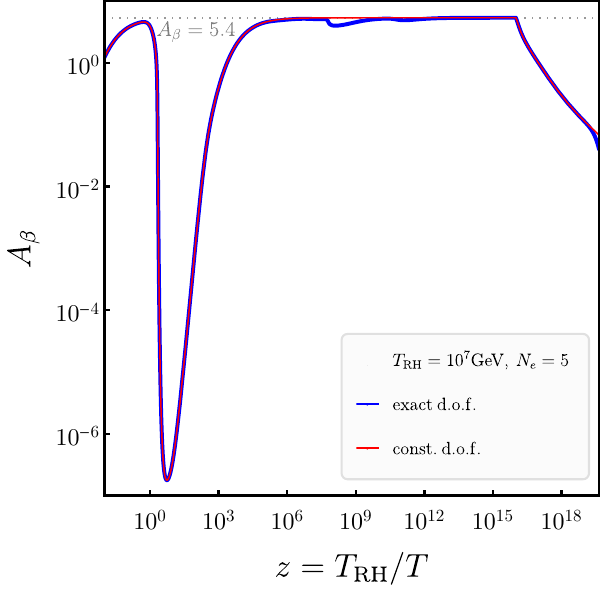}\includegraphics[scale=.75]{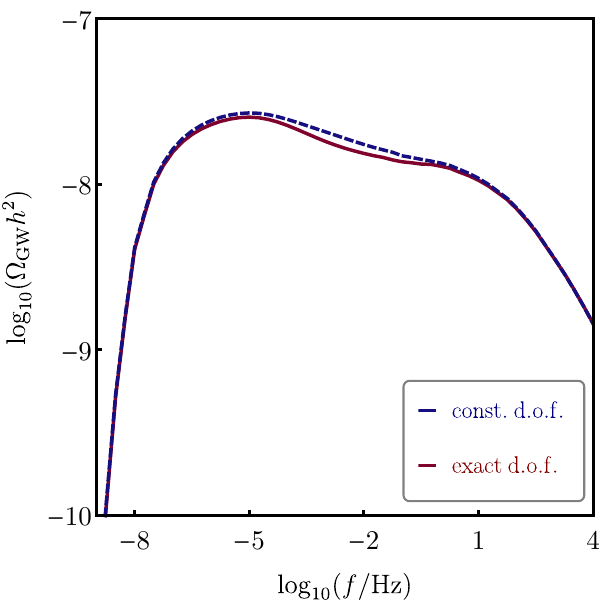}\\
\includegraphics[scale=.75]{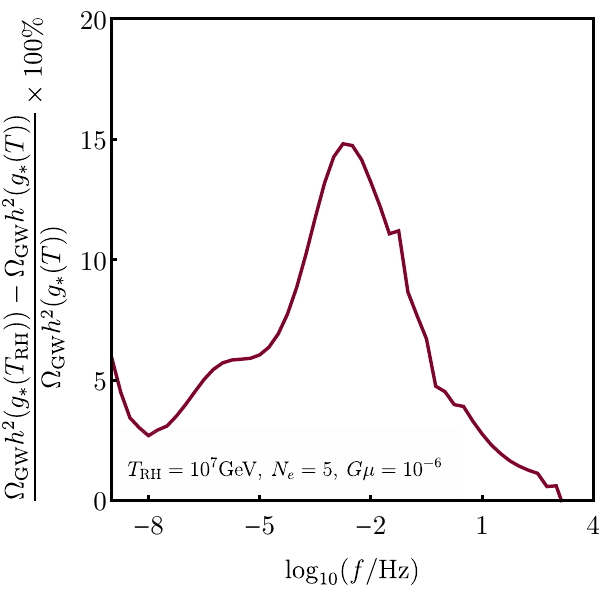}\includegraphics[scale=.75]{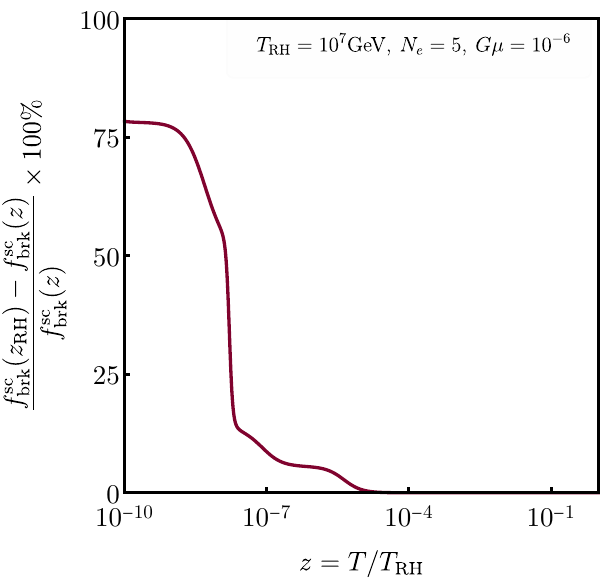}

 	\caption{\textit{Top Left}: Evolution of the loop production efficiency $A_\beta$ taking into account the exact degrees of freedom. \textit{Top Right}: Gravitational wave spectra for $G\mu=10^{-6}$, $\sqrt{\kappa}=7.9$, $T_{\rm RH}=10^7$ GeV, shown for constant (exact) degrees of freedom with dotted (solid) line. \textit{Bottom Left}: Percent level deviation of GW amplitude, and  \textit{Bottom Right}: spectral break assuming constant degrees of freedom.}\label{dof1}
 \end{figure}

\begin{figure}
\centering
\includegraphics[scale=.75]{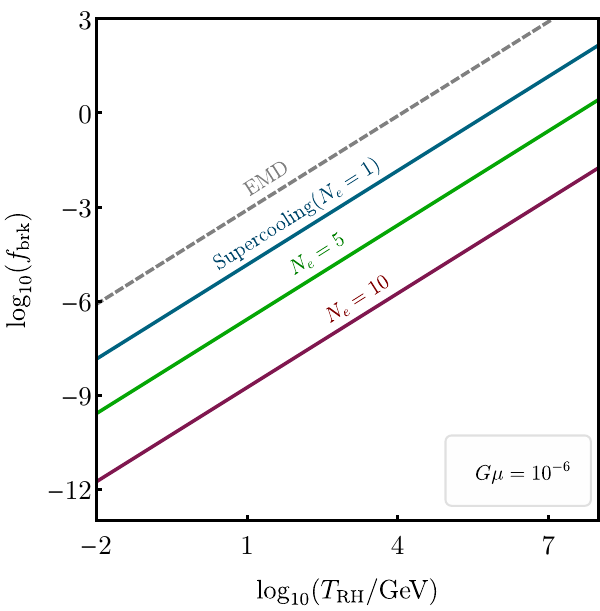}

 	\caption{A quantitative comparison of the spectral break frequency between EMD and supercooling across various number of e-folds $N_e=1,5,10$ for $G\mu=10^{-6}$.}\label{fbrkx}
 \end{figure}

Here, we present a quantitative comparison between our assumed constant and the effects of using the exact relativistic degrees of freedom \cite{Saikawa:2018rcs}. In the top left plot of Fig. \ref{dof1}, we observe that the variation in degrees of freedom during BBN leads to a significant deviation in the loop production efficiency $A_\beta$ from the constant case. This variation affects the GW amplitude and spectral break, as indicated in the other plots. While we see a maximum deviation of 15\% in the amplitude calculations, if reentry occurs during BBN, the spectral break can be significantly altered.
{\section{Analytical insight into the derivation of the spectral break frequencies, comparison between EMD and supercooling phase }\label{appB}}
During EMD, the universe expands more slowly ($a\propto t^{2/3}$) compared to radiation domination ($a\propto t^{1/2}$). Both the Hubble length $H^{-1}\propto a^{3/2}$ and correlation length $L\sim a$ grow with time, making their difference milder, thus re-entry occurs earlier. Since long strings are stretched due to the Hubble expansion, with lengths proportional to the horizon size at the end of EMD a faster growth of the Hubble scale in radiation domination causes long strings formed during EMD to re-enter the horizon over time (at $T_{\rm end}$). This triggers additional loop formation via the interaction and fragmentation of re-entered strings, causing a characteristic break at frequency $f_{\rm brk}^{\rm EMD}$ 
\begin{equation}
    f_{\rm brk}^{\rm EMD}= 0.08\:{\rm Hz}\left( \frac{T_{\rm end}}{10 {\:\rm GeV}}\right) \left( \frac{0.1\times 50\times 10^{-10}}{\alpha \Gamma G\mu}\right)^{1/2}\left( \frac{g_*(T_{\rm end})}{g_*(T_0)}\right)^{1/4} .
\end{equation}
In the case of supercooling, the universe undergoes a brief exponential expansion ($a\propto e^{Ht}$) driven by a scalar field's potential energy, which ends with a strong FOPT reheating the universe to a very high temperature $T_{\rm RH}$. The Hubble parameter remains practically constant, but the correlation length stretched exponentially since $L\sim a \propto e^{N_e}$, causing a much later re-entry. Similar to the EMD, it stretches the long strings but exponentially (network enters a frozen phase) beyond the horizon during $N_e$ e-folds. The longer the e-folds of supercooling, the stronger the stretching effect. The strength of e-folds impacts the temperature, $T_{\rm re}$, at which the long-string network re-enters the Hubble horizon, which can be obtained by setting
\begin{equation}
    L(T_{\rm re})=H^{-1}(T_{\rm re}).
\end{equation}
To characterize the configuration of a typical cosmic string network prior to thermal inflation, we begin with the assumption \cite{Guedes:2018afo,Gouttenoire:2019kij}
\begin{equation} \label{a1}
    L_i\sim 0.1 H_{\rm inf}^{-1}.
\end{equation}
During the thermal inflation phase,
\begin{equation}
L\propto a, \text{while\:} H\sim {\rm constant} \Rightarrow LH\propto a ,
\end{equation}
and by the end of inflation, it evolves to
\begin{equation}\label{a2}
(LH)_{\rm end}=(LH)_i \left(\frac{a_{\rm end}}{a_{i}}\right).
\end{equation}
After inflation ends and until the long strings re-enter the horizon, the Hubble parameter decreases as $H\propto a^{-2}$, thus
\begin{equation}\label{a3}
 LH\propto a.a^{-2}\Rightarrow(LH)_{\rm end}=\left( \frac{a_{\rm re}}{a_{\rm end}}\right).
\end{equation}
By using the fact that the energy density effectively remains conserved throughout thermal inflation, i.e. 
\begin{equation} \label{a4}
    \rho_{\rm inf}=\rho_{\rm end}=\rho_{\rm re} \left(\frac{a_{\rm re}}{a_{\rm end}} \right)^4
\end{equation}
and using \eqref{a1}, \eqref{a2}, \eqref{a3} one gets
\begin{equation} \label{reentry}
    \rho_{\rm re}^{1/4}=\frac{\rho_{\rm inf}^{1/4}}{(0.1). e^{N_e}}\Rightarrow T_{\rm re}\simeq \left(\frac{g_*(T_{\rm inf})}{g_*(T_{\rm re})}\right)^{1/4}\frac{T_{\rm RH}}{(0.1).e^{N_e}}.
\end{equation}
This expression reveals that the re-entry temperature is exponentially suppressed by the number of e-folds $N_e$ during thermal inflation compared to the scenarios involving EMD. The corresponding break frequency can be numerically approximated as
\begin{equation}\label{rebrk}
    f_{\rm brk}^{\rm sc}= 3.9\times 10^{-6} {\:\rm Hz}\left( \frac{T_{\rm re}}{10 {\:\rm GeV}}\right) \left( \frac{0.1\times 50\times 10^{-6}}{\alpha \Gamma G\mu}\right) ^{1/2}\left( \frac{g_*(T_{\rm re})}{g_*(T_0)}\right)^{1/4} ,
\end{equation}
in close agreement with earlier estimates \cite{Gouttenoire:2019kij}. By substituting Eq.\eqref{reentry} into Eq.\eqref{rebrk}, one obtains a more explicit form connecting the break frequency directly to the model parameters $T_{\rm RH}$ and $N_e$ as
\begin{equation}
    f_{\rm brk}^{\rm sc}=3.9\times 10^{-6}{\rm Hz} \left(\frac{g_*(T_{\rm RH})}{g_*(T_0)}\right)^{1/4}\left(\frac{ 0.1\times 50\times 10^{-6}}{\alpha \Gamma  G \mu}\right)^{1/2} \left(\frac{T_{\rm RH}}{\exp(N_e)} \right),
\end{equation}
assuming that there is no change in degrees of freedom between $T_{\rm inf}$ and $T_{\rm RH}$, i.e. $g_*(T_{\rm inf})=g_*(T_{\rm RH})$. This demonstrates that even if a supercooled phase transition occurs at high energy scales, the resulting spectral feature can be shifted to lower frequencies due to an extended inflationary phase. This contrasts with EMD scenarios, where horizon re-entry occurs more promptly once the Universe transitions back to radiation domination. This feature is illustrated in Fig.~\ref{fbrkx}, which shows that in the supercooling case, depending on $N_e$, the spectral break can occur at $f \sim 10^{-6}$~Hz, even for reheating temperatures as high as $T_{\rm RH} \sim 10^7$~GeV.

\section{More on the multifaceted nature of supercooling}

 \subsection{SIGWs and PBH DM in supercooling scenarios}\label{appc1}
 During a first-order phase transition, the universe can give rise to black holes through a variety of mechanisms \cite{Liu:2021svg,Kawana:2022olo,Lewicki:2023ioy,Kierkla:2025vwp}. One particularly intriguing pathway involves the statistical variations in bubble nucleation histories across different regions of space. Because these bubbles don't all form and merge simultaneously, some regions—known as causal patches —complete their transitions later than others. These late-blooming patches experience extended periods dominated by vacuum energy, allowing matter to accumulate to such an extent that the resulting overdensities can collapse into PBHs \cite{Gouttenoire:2023naa}\footnote{However, a recent study  \cite{Franciolini:2025ztf} highlights the importance of gauge fixing and points out that the final production of PBHs and SIGWs can be significantly impacted due to a mismatch between the gauge choices used in the computation.}. The likelihood of this occurring within a given Hubble-sized patch, depending on the underlying nucleation dynamics, can be characterized by the probability
 \begin{equation}
     \mathcal{P}_{\rm coll}=\exp\left[ -a\left(\frac{\beta}{H_\star} \right)^b (1+\delta_c)^{c\beta/H_\star}\right],
 \end{equation}
 where $a=0.5646$, $b=1.266$, $c=0.6639$, and $\delta_c\sim 0.45$. The present-day fraction $f_{\rm PBH}$ of DM in the form of PBHs is given by
 \begin{equation}
     f_{\rm PBH}=\left(\frac{\mathcal{P}_{\rm coll}}{2.2\times 10^{-11}}\right) \left(\frac{T_{V}}{140 \:\rm GeV}\right).
 \end{equation}
  Meanwhile, the mass of these PBHs is given by the energy content within the sound horizon at the time of their gravitational collapse \cite{Escriva:2021pmf}
 \begin{equation}
     M_{\rm PBH}=M_{\odot} \left(\frac{20}{g_*(T_{V})}\right)^{1/2} \left( \frac{0.14 \rm GeV}{T_{V}}\right)^{2}.
 \end{equation}
 Here we are specifically interested on the case where these PBHs can explain $100\%$ of DM, i.e. $f_{\rm PBH}=1$. However, such a scenario remains viable only within a narrow and tightly constrained mass window, as dictated by current astrophysical and cosmological bounds \cite{Cirelli:2024ssz}
 \begin{equation}
     10^{16} M_\odot\lesssim M_{\rm PBH} \lesssim 3\times 10^{-12} M_\odot,
 \end{equation}
 which roughly corresponds to
 \begin{equation}\label{ConsRH}
     50 \text{ TeV}\lesssim T_{\rm RH}(\sim T_V) \lesssim 10 \text{ PeV}.
 \end{equation}
 
  \begin{figure}
\centering

\includegraphics[scale=.75]{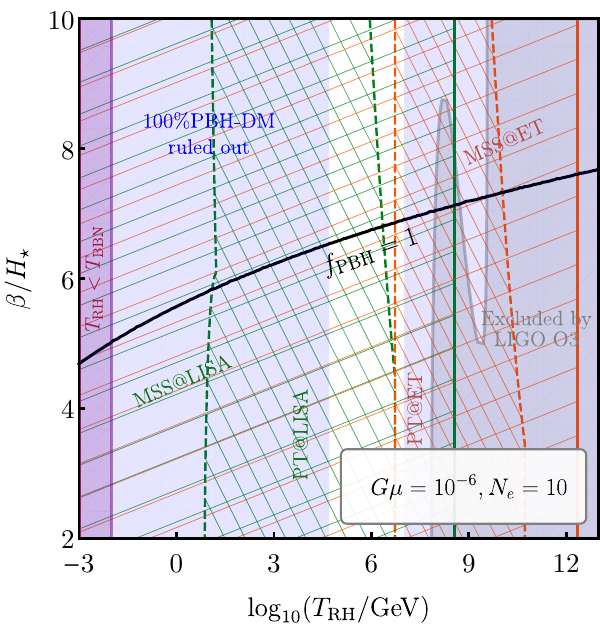}\includegraphics[scale=.75]{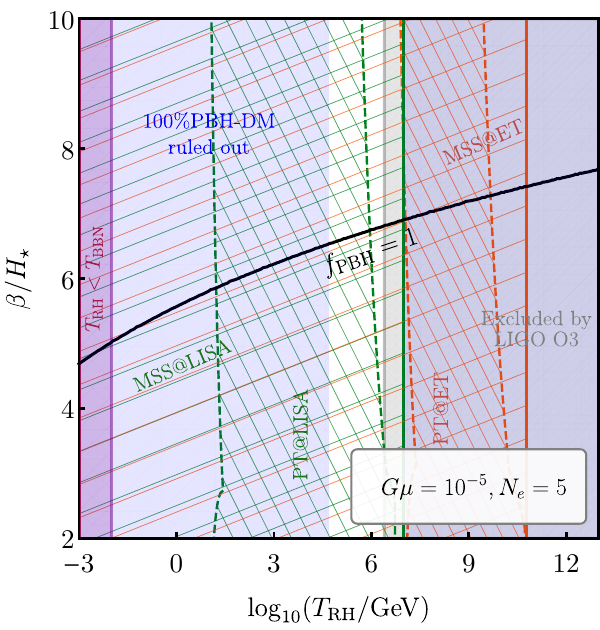}

 	\caption{\textit{Left}: Region of parameter space where the modified gravitational wave signal, featuring both cosmic string and FOPT contributions, is detectable with signal-to-noise ratio $\mathrm{SNR} \gtrsim 10$. The intersection of the CS and FOPT regions represents the ``smoking gun" parameter space where FOPT dominates. The analysis separates the modified cosmic string tail (feature beyond $f_{\rm brk}^{\rm sc}$) from the FOPT signal, considering only portions of the spectrum where the amplitude deviates by at least 10\% from that of a standard string spectrum. For this plot, we fix the PTA favored string tension at $G\mu = 10^{-6}$ and the duration of supercooling at $N_e = 10$. The black contour traces the boundary where PBHs account for all DM, while the shaded blue region is ruled out by astrophysical and cosmological limits, as described by \eqref{ConsRH} \textit{Right}: Similar parameter scan, but with $G\mu=10^{-5}$ and $N_e=5$.}\label{PBH_DM}
 \end{figure}
 Fig. \ref{PBH_DM} illustrates the impact of the PTA fit by metastable strings at low frequencies while the high-frequency region is predominantly influenced by GWs from supercooled phase transition, as naturally motivated from PBH-DM scenario which requires a sufficiently small $\beta/H_\star\sim 6-8$ for $f_{\rm PBH}=1$, as indicated by the black curve. What makes this setup particularly compelling is that, across much of the viable parameter space defined by \eqref{ConsRH}, there's a tantalising possibility: simultaneous observation of the change of spectral shape of cosmic string GW spectrum and a distinct FOPT-induced GW peak layered atop the background. However, despite the PTA's preference for string tension $G\mu \gtrsim 10^{-5}$, the lack of any SGWB detection in the LIGO O3 run imposes severe constraints on the PBH-DM landscape.\\
    
\begin{figure}
\centering

\includegraphics[scale=.75]{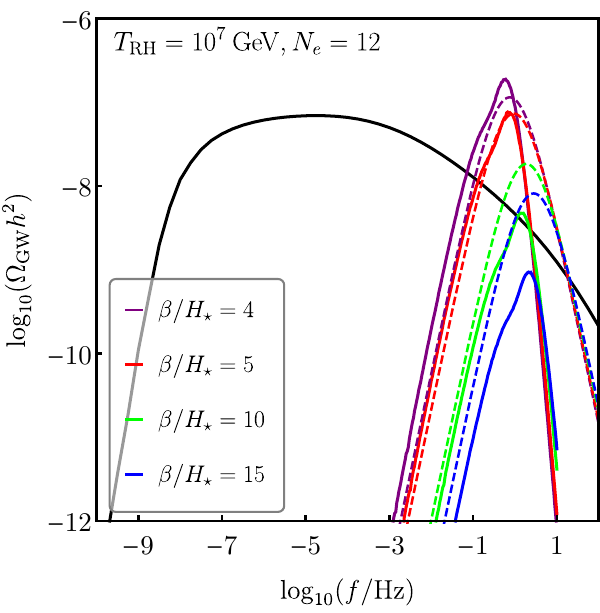} \includegraphics[scale=.75]{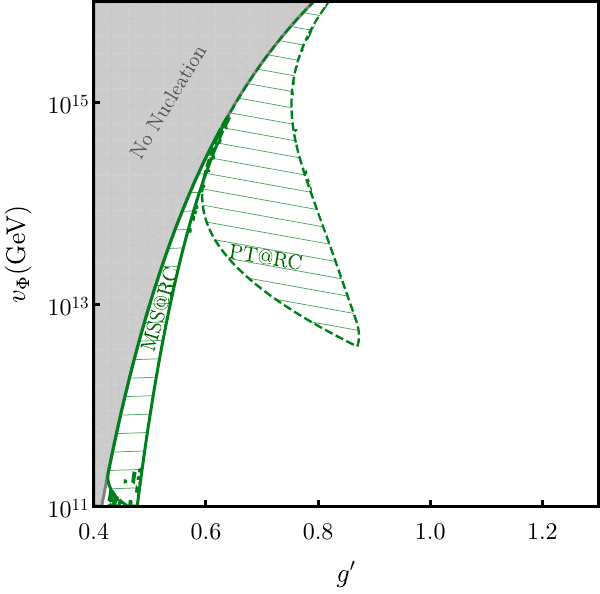}\\

 	\caption{\textit{Left}: The possible hybrid GW spectrum combining contributions from metastable cosmic strings (solid black), first-order phase transitions (dashed), and curvature perturbations (solid). This composite signal emerges naturally within the PBH–DM framework driven by supercooled FOPT dynamics. \textit{Right}: Allowed parameter space in the minimal conformal $B\!-\!L$ model that leaves detectable signatures in resonant cavity (RC) experiments \cite{Herman:2022fau}. In the region labeled \textbf{MSS@RC}, the modified cosmic string tail (feature beyond $f_{\rm brk}^{\rm sc}$) due to supercooling is detectable, while in the region labeled \textbf{PT@RC}, the signal is dominated by the first-order phase transition over the cosmic string background.}\label{SIGW}
 \end{figure}

 Another interesting feature of scenarios favoring $f_{\rm PBH}=1$, is the emergence of an additional low-frequency gravitational wave signal of scalar-induced gravitational waves (SIGW) arising from enhanced curvature perturbations during the supercooling epoch \cite{Elor:2023xbz,Gouttenoire:2025wxc}. When $\beta/H_\star$ becomes sufficiently small, these scalar modes can significantly source gravitational waves, with the curvature power spectrum given by:
 \begin{equation}
     \mathcal{P}_{\zeta}(k)=\left( \frac{\alpha_{\rm PT}}{1+\alpha_{\rm PT}}\right)^2 \left( \frac{H_\star}{\beta}\right)^2 \frac{0.9(k_p/\beta)^3}{\left( 0.24+(k_p/\beta)^2\right)^3} \Theta(H_\star-k_p),
 \end{equation}
where $k_p=k/a_*$. These perturbations evolve into today’s SGWB with a present-day energy density expressed as \cite{Kohri:2018awv,Espinosa:2018eve,Inomata:2019yww}
 \begin{equation}
     \Omega_{\rm GW}^{\rm SIGW}= c_g \Omega_{r,0} \int_{0}^{\infty}dv\int_{|1-v|}^{1+v}du \left( \frac{4v^2-(1+v^2-u^2)^2}{4uv}\right)^2 \bar{\mathcal{I}}_{u,v}^2 \mathcal{P}_{\zeta}(kv) \mathcal{P}_{\zeta}(ku),
 \end{equation}
 where the transfer function can be given by \cite{Kohri:2018awv}
 \begin{equation}
     \bar{\mathcal{I}}_{u,v}^2=\frac{1}{2}\left(\frac{3(u^2+v^2-3)^2}{4u^3v^3}\right)^2\left[ \left(-4uv+(u^2+v^2-3)\log\bigg|\frac{3-(u+v)^2}{3-(u-v)^2}\bigg|\right)^2+\pi^2(u^2+v^2-3)^2\Theta(u+v-\sqrt{3}) \right]
 \end{equation}
  In fact, when $\beta/H_\star \lesssim 5$, the SIGW contribution can become dominant as shown in Fig. \ref{SIGW} (left).

\subsection{An explicit realisation: scFOPT-triggered string formation} \label{appc2}
We consider a well-studied minimal conformal $(B-L)$ extension \cite{Iso:2009nw,Iso:2009ss} of the Standard Model, where the renormalized, loop-corrected form of the vacuum energy can be expressed as,
\begin{equation}\label{blV}
    \rho_V=\frac{\beta_{\lambda_{\rm eff}}v_\Phi^4}{16}
\end{equation}
with
\begin{equation}
    \beta_{\lambda_{\rm eff}}\simeq \frac{6 {g^\prime}^4}{\pi^2}
\end{equation}
Using \eqref{blV}, one can get a model-dependent form of the reheating temperature,
\begin{equation}
    T_{\rm RH}=\frac{1}{\pi}\left(\frac{45}{4g_*(T_{\rm RH})}\right)^{1/4} g^\prime v_\Phi
\end{equation}
and using \eqref{tension} one can directly relate the $T_{\rm RH}$ with $G\mu$ as
\begin{equation}
   T_{\rm RH}=3.83\times 10^{15} \: {\rm GeV} \left( \frac{108}{g_*(T_{\rm RH})}\right)^{1/4}  \left( \frac{g^\prime}{0.5}\right)  \left( \frac{G\mu}{10^{-6}}\right)^{1/2}.\label{cor}
\end{equation}
where we can safely ignore the effect of $h(\lambda,g^\prime)$ since $g^\prime>0.4$ for $v_\Phi>10^{11}$ GeV, according to the nucleation condition.\\




In our model-specific study, we apply the analytical thick wall approximation, allowing us to approximate the Euclidean action as follows \cite{Baldes:2021aph}
\begin{equation}
    \frac{S_3}{T}\simeq \frac{A}{\log(M/T)}
\end{equation}
with
\begin{equation}
    A=1.11P/\beta_{\lambda_{\rm eff}},\: M= 4.83\beta_{\lambda_{\rm eff}}^{1/2} v_\Phi/P,\:{\rm and}\: P=\sqrt{226.2 {g^\prime}^2
-96  {g^\prime}^3}.\end{equation}
According to standard tunnelling formalism, the space-time density of bubble nucleation rate at finite temperatures is \cite{Linde:1981zj}
\begin{equation}
    \gamma\sim T^4 \left( \frac{S_3}{2\pi T}\right)^{3/2} e^{-S_3/T}.
\end{equation}
Using the relation $\gamma(T_n)=H^4(T_n)=H_V^4$, we can derive an approximate expression for the nucleation temperature,
\begin{equation}
    T_n \simeq \sqrt{MH_V}\exp \left(\frac{1}{2} \sqrt{-A+\log^2(M/H_V)} \right)
\end{equation}
with the condition for no nucleation given by $T_n^{\rm min}=\sqrt{MH_V}$. 
The PT parameters can be expressed as
\begin{equation}
    \alpha_{\rm PT}\sim \frac{\rho_V}{\rho_{\rm rad}(T_n)}=\left(\frac{T_{\rm RH}}{T_n}\right)^4
\end{equation}
and
\begin{equation}
    \frac{\beta}{H_\star} = -\frac{\partial \log \gamma}{\partial \log T}\bigg|_{T_n}\simeq -4 +\frac{A}{\log^2(M/T_n)}.
\end{equation}
Therefore, we can also approximate $N_e$ in terms of model parameters,
\begin{equation}
    N_e=\log\left( \frac{T_{\rm RH}}{T_n}\right).
\end{equation}
Consequently, in this scenario, we can express the spectral break frequency in terms of the model parameters $g^\prime$ and $v_{\Phi}$.\\

We focus on two key spectral features: the modified cosmic string tail beyond $f_{\rm brk}^{\rm sc}$ and the characteristic peak from the first-order phase transition, which dominates over the string background, both of which become more pronounced at smaller $g^\prime$, as illustrated in the right panel of Fig. \ref{SIGW}. This is because a lower $g^\prime$ allows for increased supercooling while the reheating temperature decreases. However, nucleation ceases below a critical $g^\prime$, driving the universe into eternal inflation. Additionally, increasing the symmetry-breaking scale $v_\Phi$ raises the string tension and shifts the break to lower frequencies, but at the same time elevates the reheating temperature, partly compensating the effect. In this minimal setup, the lower bound on $g^\prime$ significantly restricts the parameter space, pushing observable features to high frequencies. Nonetheless, more general constructions may relax these constraints, allowing stronger supercooling and delayed reheating, and thereby shifting the signals into experimentally accessible frequency ranges.
 
 \bibliography{main.bib}
\end{document}